\documentclass[sigconf,screen]{acmart}

\AtBeginDocument{%
  }

\author{Roberto Pietrantuono}
\orcid{0000-0003-2449-1724} 
\email{roberto.pietrantuono@unina.it}
\affiliation{\institution{University of Naples Federico II} \city{Naples} \country{Italy}}

\author{Luca Giamattei}
\orcid{0000-0003-3767-4036}
\email{luca.giamattei@unina.it}
\affiliation{\institution{University of Naples Federico II} \city{Naples} \country{Italy}}

\author{Stefano Russo}
\orcid{0000-0002-8747-3446} 
\email{stefano.russo@unina.it}
\affiliation{\institution{University of Naples Federico II} \city{Naples} \country{Italy}}

\author{Julien Siebert}
\orcid{0000-0002-7696-0046}
\email{julien.siebert@iese.fraunhofer.de}
\affiliation{\institution{Fraunhofer IESE} \city{Kaiserslautern} \country{Germany}}

\author{Neil Walkinshaw}
\orcid{0000-0003-2134-6548}
\email{n.walkinshaw@sheffield.ac.uk}
\affiliation{\institution{University of Sheffield} \city{Sheffield} \country{England}}

\usepackage{xcolor}
\usepackage{soul}
\usepackage{microtype}
\usepackage{booktabs}
\usepackage{enumitem}
\usepackage{tikz}
\usetikzlibrary{positioning}

\usetikzlibrary{arrows.meta,positioning,calc,fit}

\begin{document}

\setcopyright{acmlicensed}
\acmYear{2026}\copyrightyear{2026}
\setcopyright{cc}
\setcctype[4.0]{by}

\title{Causal Software Engineering: A Vision and Roadmap}

\begin{abstract}

Software engineering increasingly involves making high-stakes decisions under uncertainty, using signals from code, field data, and socio-technical processes. Recent AI-driven support (e.g., anomaly detection, predictive analytics, AIOps, as well as LLM-based agents) has amplified engineers’ ability to detect patterns and synthesize content and recommendations, but many critical questions are interventional or counterfactual: \textit{What is the expected impact of changing a load-balancing strategy? Would an outage have been avoided under a different release plan?} Correlational models answer "what tends to co-occur"; they struggle to answer "what would happen if we act." 
We propose \textbf{Causal Software Engineering (CSE)} as a future paradigm in which causal models and causal reasoning systematically inform activities across the software lifecycle, augmenting existing practices with explicit assumptions, uncertainty-aware effect estimates, and counterfactual diagnosis. We outline (i) a causal-first workflow view spanning development and operations, (ii) a staged roadmap for tools and organizational adoption, and (iii) an evaluation and benchmark agenda for measuring progress.

\end{abstract}



\maketitle

\section{Beyond correlation}
\label{sec:motivation}

Modern software systems are built and operated through a stream of decisions: what design alternatives to implement, what and how to test, which change to release, what to monitor, how to respond when behavior drifts. Teams now have unprecedented observability and automation, and increasingly rely on AI-enabled assistance -- from anomaly detection and forecasting to AIOps triage and LLM-based agents drafting explanations, patches, and operational playbooks. Yet, these tools often provide \emph{plausible narratives} or \emph{associational signals} rather than decision-grade answers. Many practical questions are not about association but about \emph{intervention}: what will happen if we change a timeout, flip a feature flag, re-route traffic, reconfigure and redeploy, or alter a release strategy? Others are \emph{counterfactual}: would the incident still have happened if the roll-out had been slower, if a guardrail had fired earlier, or if a configuration had been set differently? Such questions require reasoning about causes, confounders, and uncertainty, not only about co-occurring patterns in historical data or plausible ``interpolations'' from them.

\emph{Example.} A microservice team introduces a new client-side retry policy to reduce tail latency and ships it in a release. In an hour, latency drops and the AIOps dashboard attributes the improvement to the deployment, suggesting the change ``fixed latency.'' The on-call engineer plans to apply the same pattern to other services. However, during the same window a separate auto-scaling adjustment increased replicas in a downstream dependency (and a traffic shift moved requests away from a ``hot'' region), both of which can reduce queueing delays. The improvement is therefore ambiguous: it could be the retry-policy change, the scaling change, the traffic shift, or their interaction. When the team later repeats the roll-out without the same scaling conditions, tail latency regresses and an incident occurs, because the original causal driver was misidentified.

We envisage a new paradigm -- \emph{Causal Software Engineering (CSE)} -- where causal models and causal reasoning are integrated into everyday software engineering (SE) artifacts and workflows. CSE does not replace existing SE practices; it augments them with explicit assumptions, uncertainty-qualified \textit{causal effect} estimates, and counterfactual diagnosis. In CSE, the release above is treated as an \emph{intervention}, denoted with the \textit{do-operator} ($do(X=x)$), and the concurrent autoscaling change and traffic/region mix are recorded as potential \emph{confounders} ($Z$) in an intervention log because they affect latency and may co-vary with the release, creating spurious correlations in observational data. A CSE engineer (or tool) estimates the causal effect of $X$ on tail latency, adjusting for $Z$ (and other confounders such as time-of-day, cache warmup, and parallel configuration changes).
Estimating the \textit{free-from-confounding} causal effect (with uncertainty bounds), the system reports a decision-ready estimate and recommends a concrete next step:
(i)~\emph{proceed with roll-out} (or increase the canary percentage) when the impact is beneficial;
(ii)~\emph{proceed, but avoid further tuning} when the improvement is real but modest;
or (iii)~\emph{target the likely cause} by applying the specific code/config change elsewhere (or tuning the next parameter) when the model attributes the effect to a particular factor.

\section{Causality-centered software engineering}
\label{sec:vision}

Although correlation-based models -- including many of today's ML systems and LLM-based agents \cite{Hou2024,Fan2023,Wang2023} -- can be large and sophisticated, they inherently lack causal reasoning on their own; it requires structure beyond pattern learning. The key enabler remains data, but the crucial step is to \emph{formulate} SE questions causally: what is the intervention, what outcome is affected, and what assumptions are needed to identify an effect.

\textbf{Thesis.} The next wave of SE tooling will treat changes (code, configuration, design/testing choices, deployment and remediation actions) as causal \emph{interventions} on a system-in-environment, moving engineers from \textit{post hoc} correlation hunting to \textit{intervention-aware} decisions with \emph{explicit assumptions}, quantified uncertainty, and auditable \emph{causal claims}.

\textit{From correlation to causation.}
Many current tools learn from historical co-occurrences: ``when metric $Y$ is high, event $E$ often appears,'' or ``after change $X$, $Y$ tends to move.'' Such associations aid detection and triage but do not justify actions. The core CSE question is different: \emph{what will happen if we do $X$?} \cite{Pearl2009}
CSE reasons about interventional distributions $P(Y|do(X=x))$, i.e., the confounder-free causal effect of setting $X$ on outcome $Y$ (e.g., latency, crashes, fairness), and about counterfactuals (what would have happened under a different $x$). This matters because software evolves and environments shift; spurious correlations that held yesterday can break tomorrow when hidden factors change. CSE enables queries like:
\begin{itemize}[leftmargin=*,nosep]
\item \textbf{Interventional (what-if):} If we set $X = x$ (e.g., replicas, feature flag, timeout), how will $Y$ (e.g., tail latency, crash rate, fairness metric) change?
\item \textbf{Counterfactual (what-would-have-happened):} Would the incident have occurred if $X$ had been different?
\end{itemize}

\textit{What is a causal model (in SE terms)?}
A \emph{causal model} compactly represents how relevant factors causally influence each other. In practice, CSE uses \emph{task-specific} models tied to concrete decisions (e.g., ``did this release reduce p99 latency?'' or ``which configuration change caused the regression?''). One convenient form is a directed acyclic graph: nodes are observable or controllable variables (release version, retry policy, replicas, cache warm-up, region mix) and edges encode hypothesized cause--effect relations, with data-parameterized equations quantifying their strength. The model can be defined by an expert from domain knowledge, requirements, and design documents (encoding assumptions, e.g., about forbidden or required relations); learned from data like any ML model; or built via a hybrid approach \cite{wang24_cd}. With such a model, tools answer ``what-if'' questions by estimating the effect of $X$ while accounting for other factors.

\textit{Why ``explicit assumptions'' are different from today.}
All predictive tools rely on assumptions, but many remain implicit: which variables matter, which are merely correlated, and which hidden factors could create spurious conclusions. CSE makes these \emph{explicit and reviewable}:
\begin{itemize}[leftmargin=*,nosep]
\item \textbf{Potential confounders.} A modeler specifies that traffic/region mix or autoscaling changes may influence both the likelihood of deploying a mitigation and the observed latency; the data is then used to estimate effects under this adjustment and test stability.

\item \textbf{Allowed or forbidden causal links.} The modeler may \emph{forbid} an arc \texttt{Latency} $\rightarrow$ \texttt{Deployed\_Version} (latency spikes do not change which version is deployed) or \emph{require} \texttt{Region\_Mix} $\rightarrow$ \texttt{Latency} (traffic composition affects latency); the causal analysis is then constrained to models consistent with these priors.

\item \textbf{When a causal claim is justified.} The model states what must be measured (and what must \emph{not} be simultaneously changing) for an effect estimate to be identifiable.
\end{itemize}

By externalizing assumptions, CSE lets engineers inspect, contest, revise, and encode them --- similarly to reviewing code, ADRs, and runbooks. Unlike black-box correlations, when assumptions do not hold, CSE surfaces which one failed, e.g., a hypothesized confounder changes with the intervention, a forbidden/required link is contradicted, or the identification conditions (what must be measured and what must remain stable) are not satisfied.

\textit{Revisiting the running example.}
Returning to the microservice rollout in Section~\ref{sec:motivation}, CSE differs from correlational tooling by making \emph{causal assumptions} explicit and checkable. First, the intervention log records what is being changed ($X$) and concurrent confounders ($Z$, e.g., autoscaling, traffic/region mix), rather than treating them as incidental context. Second, the causal design spec constrains the structure via domain knowledge (e.g., forbids \texttt{Latency}$\rightarrow$\texttt{Deployed\_ Version}, requires \texttt{Region\_Mix}$\rightarrow$\texttt{Latency}). Third, the spec encodes \emph{identifiability conditions}: what must be measured (e.g., scaling events, traffic mix) and what must remain stable (e.g., no simultaneous platform-wide changes) for the effect estimate to be justified. When conditions hold, the runtime model reports an estimate with uncertainty to guide rollout; otherwise, it reports the failed assumption and recommends a next step (canary/AB test, isolate the intervention, or revise the model).

\begin{table*}[t!]
\caption{Examples of causal questions across the lifecycle and their primary CSE artifact.}
\label{tab:lifecycle}
\footnotesize
\begin{tabular}{@{}p{0.08\linewidth}p{0.52\linewidth}p{0.36\linewidth}@{}}
\toprule
\textbf{Stage} & \textbf{Causal question (example)} & \textbf{Output (CSE artifact)}\\
\midrule
Requirements &
If we change policy $X$ (e.g., ranking algorithm), what is the effect on user harm or fairness $Y$? &
\textbf{Causal design spec} (outcomes, confounders, admissible interventions)\\
Architecture &
If we split service $A$ or change sync/async topology, what is the effect on tail latency and incident rate? &
\textbf{Causal design spec} (architecture interventions, causal assumptions)\\
Testing &
Which scenario interventions maximize discovery of safety-critical failures; which factors truly cause violations? &
\textbf{Live causal model} (factor model and uncertainty)\\
Debugging &
Would the failure have occurred if configuration $X$ were different? &
\textbf{Live causal model} (counterfactual RCA report)\\
Operations &
Which remediation action caused improvement versus correlated recovery noise? &
\textbf{Intervention log} (with effect estimate plus uncertainty)\\
\bottomrule
\end{tabular}
\end{table*}

\textit{Causal workflow across the lifecycle.}
To realize this vision, we propose lightweight \textbf{CSE artifacts} that fit into existing pipelines:
\begin{enumerate}[leftmargin=*, topsep=3pt, itemsep=0pt]
\item \textbf{Causal design specs.} Small, versioned causal assumptions attached to requirements, designs, and ADRs, capturing (i)~key variables and outcomes, (ii)~candidate confounders, and (iii)~admissible interventions. \\
\emph{Example:} \textit{Decision}: enable a client-side retry policy ($X$) to reduce tail latency ($Y$) without increasing error rate.
\textit{Confounders}: autoscaling events and replica count; traffic/region mix; request rate and load; time-of-day; concurrent configuration changes (e.g., timeouts, circuit breakers).
\textit{Admissible interventions}: toggle retry policy via feature flag; adjust retry count and backoff; revert to previous policy.

\item \textbf{Intervention logs.} Structured records of changes (deployments, feature toggles, mitigations) with intended causal outcomes, context, and uncertainty, connecting CI/CD actions to operational reality: what changed, when, where, what else changed, and which outcomes should move. \\
\emph{Example:} Release v2.3 enabled \texttt{retry\_policy=v2} for a 10\% canary in \texttt{eu-west} (\textit{intervention}); during the same window, a downstream service scaled from 20 to 40 replicas and 12\% of traffic shifted from \texttt{eu-west} (\textit{confounders}); \textit{objective}: reduce p99 latency by 10\% without increasing error rate.

\item \textbf{Living Causal Model.}
The design spec defines the causal structure; the runtime model instantiates it with live telemetry and (where possible) experimental data, maintaining up-to-date effect estimates and uncertainty. The living model may be partial, focusing on a service boundary, an SLO, or an incident class. Its role is \emph{decision support under interventions}: what-if queries and counterfactual diagnosis with uncertainty and refutation checks.
\end{enumerate}

Table~\ref{tab:lifecycle} sketches how causal questions arise throughout SE; each stage produces a lightweight artifact (spec/log/living model update) that can be reviewed, versioned, and audited like code

\section{Roadmap}
\label{sec:roadmap}
\begin{figure*}[t]
\centering
\footnotesize

\begin{tikzpicture}[
  font=\footnotesize,
  box/.style={draw, rounded corners=2pt, inner sep=2pt, align=left},
  coltitle/.style={font=\bfseries},
  lane/.style={minimum height=7mm},
  arrow/.style={-Latex, thick}
]

\def\xA{0}
\def\xC{6.8}
\def\xD{13}

\def\yone{2.7}
\def\ytwo{1.1}
\def\ythree{-0.3}
\def\yfour{-1.6}

\node[coltitle] (tA) at (\xA,3.6) {Achievements (``We are here'')};
\node[coltitle] (tC) at (\xC,3.6) {Challenges (research directions)};
\node[coltitle] (tD) at (\xD,3.6) {Dreams (asymptotic goals)};

\node[box, lane, text width=5.5cm] (a1) at (\xA,\yone) {\textbf{Route 1: Causal observability}\\Causal fault localization and RCA graphs; causal dependency models from logs and architectures; causal models of design, process, quality trade-offs};
\node[box, lane, text width=5.5cm] (a2) at (\xA,\ytwo) {\textbf{Route 2: Intervenability-by-design}\\Retrospective causal impact of refactoring/patches; CI/CD tools for staged change (flags, canaries, rollbacks); quasi-experimental analysis of architectural/design changes};
\node[box, lane, text width=5.5cm] (a3) at (\xA,\ythree) {\textbf{Route 3: Counterfactual assurance}\\Counterfactual debugging and execution comparison; value-based counterfactual fault localization; counterfactual testing and scenario analysis};
\node[box, lane, text width=5.5cm] (a4) at (\xA,\yfour) {\textbf{Route 4: Governance \& alignment}\\ Safety cases; compliance processes; \\model risk management};

\node[box, text width=6cm] (c1) at (\xC,\yone) {\textbf{Causal observability}\\
\strut $\bullet$ confounder-aware models over monitoring data\\
$\bullet$ stable causal graphs under evolution\\
$\bullet$ causal variable engineering from artifacts};

\node[box, text width=6cm] (c2) at (\xC,\ytwo) {\textbf{Intervenability-by-design}\\
\strut $\bullet$ safe experiments (canary/AB), identification\\
$\bullet$ uncertainty tolerance for interventions\\
$\bullet$ causal impact of configuration/deployment changes};

\node[box, text width=6cm] (c3) at (\xC,\ythree) {\textbf{Counterfactual assurance}\\
\strut $\bullet$ counterfactual RCA at scale\\
$\bullet$ causal testing (factors \& interventions)\\
$\bullet$ causal refutation checks as standard};

\node[box, text width=6cm] (c4) at (\xC,\yfour) {\textbf{Governance \& alignment}\\
\strut $\bullet$ auditable causal claims for safety/fairness\\
$\bullet$ human-in-the-loop causal modeling, tooling\\
$\bullet$ LLM-based agents constrained by causal consistency};

\node[box, text width=4.5cm] (d1) at (\xD,\yone) {\textbf{Causal diagnostic explanations}\\ Reliable what-if and why \\answers grounded in causal models};
\node[box, text width=4.5cm] (d2) at (\xD,\ytwo) {\textbf{Intervention planning}\\ Continuous causal optimization\\ of performance, reliability, and cost};
\node[box, text width=4.5cm] (d3) at (\xD,\ythree) {\textbf{Counterfactual engineering}\\ Routine prevention planning: \\"what would have avoided this?"};
\node[box, text width=4.5cm] (d4) at (\xD,\yfour) {\textbf{Causal copilots}\\  Governed, uncertainty-aware assistants for safe action; Evidence-backed, reviewable causal arguments for assurance/compliance};
\draw[arrow] (a1.east) -- (c1.west);
\draw[arrow] (c1.east) -- (d1.west);

\draw[arrow] (a2.east) -- (c2.west);
\draw[arrow] (c2.east) -- (d2.west);

\draw[arrow] (a3.east) -- (c3.west);
\draw[arrow] (c3.east) -- (d3.west);

\draw[arrow] (a4.east) -- (c4.west);
\draw[arrow] (c4.east) -- (d4.west);

\draw[densely dashed] (3.3,3.6) -- (3.3,-2.1);
\draw[densely dashed] (10.25,3.6) -- (10.25,-2.1);

\end{tikzpicture}

\vspace{0.3em}

\begin{tikzpicture}[
  font=\small,
  crl/.style={draw, rounded corners=2pt, inner sep=2pt, align=left},
  arrow/.style={-Latex, thick}
]
\node[crl, text width=2.4cm] (crl0) { \textbf{CRL-0}\\ Correlation-based };
\node[crl, text width=2.55cm, right=3mm of crl0] (crl1) { \textbf{CRL-1}\\ Causal observability };
\node[crl, text width=2.55cm, right=3mm of crl1] (crl2) { \textbf{CRL-2} Intervenability by design };
\node[crl, text width=2.55cm, right=3mm of crl2] (crl3) { \textbf{CRL-3}\\ Counterf. assurance };
\node[crl, text width=2.55cm, right=3mm of crl3] (crl4) { \textbf{CRL-4}\\ Causal copilots };
\node[crl, text width=2.55cm, right=3mm of crl4] (crl5) { \textbf{CRL-5}\\ Causal certification };

\draw[arrow] (crl0.east) -- (crl1.west);
\draw[arrow] (crl1.east) -- (crl2.west);
\draw[arrow] (crl2.east) -- (crl3.west);
\draw[arrow] (crl3.east) -- (crl4.west);
\draw[arrow] (crl4.east) -- (crl5.west);
\end{tikzpicture}

\caption{A roadmap for CSE
along four co-evolving routes.
\emph{Bottom:} Causal Readiness Levels (CRLs) for tools and organizations.}
\label{fig:roadmap}
\end{figure*}
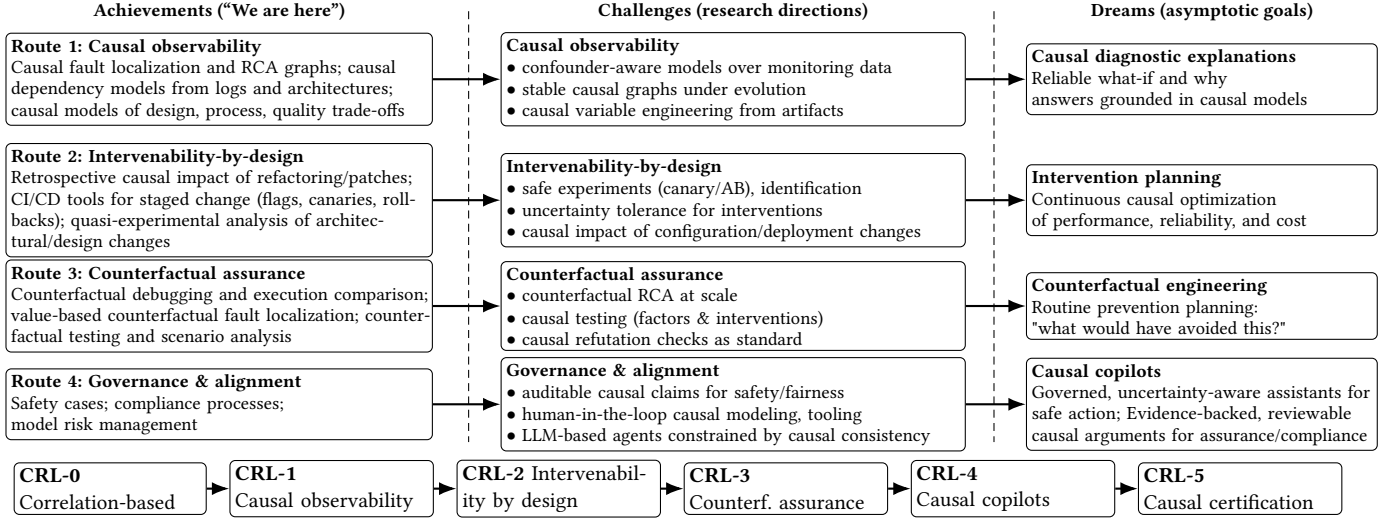

To keep CSE actionable, we present a roadmap inspired by the ``achievements--challenges--dreams'' structure of Bertolino \textit{et al.}'s software testing roadmap \cite{Bertolino2007}, identifying (i)~what we already have (achievements), (ii)~central challenges that enable progress, and (iii)~long-term ``dreams'' as asymptotic goals. Figure~\ref{fig:roadmap} organizes challenges along \emph{four routes} spanning key lifecycle activities; we interpret progress toward CSE as climbing \emph{Causal Readiness Levels (CRLs)}.

\textbf{{Route 1} (R1): Causal observability (CRL-1: \textbf{\textit{observe}}).}
R1 extends traditional observability by augmenting signals engineers collect (metrics, logs, traces, dependencies, development artifacts) with \emph{explicit causal structure}. Work in fault localization, root-cause analysis \cite{Ikram22,Johnson20,Chen19}, and modeling architectural/design decisions, process, and quality trade-offs \cite{Iqbal22,Dubslaff22} already builds causal graphs or dependency models for attribution. These achievements mark ``where we are'': causal structure is increasingly extracted from operational/process data, but is often used for retrospective ranking (e.g., root causes) rather than robust causal claims, and can be unstable under system evolution \cite{Hulse2025}. Key challenges (Fig.~\ref{fig:roadmap}): \textit{confounder-aware models over monitoring data}; \textit{causal variable engineering from artifacts}; and \textit{stable causal graphs under evolution} with drift/assumption checks. If met, R1 enables \textbf{causal explanations}: reliable why/what-if answers grounded in explicit causal structure.

\textbf{Route 2 (R2): Intervenability-by-design (CRL-2: \textbf{\textit{intervene}}).}
Teams already have strong change-control mechanisms -- CI/CD, feature flags, canaries, staged rollouts, rapid rollback -- that make interventions feasible in production, but rarely frame them as explicit \textit{causal} interventions. Causal analysis is therefore often still \emph{post hoc} \cite{Giamattei2025,Siebert23}: a change is shipped, metrics move, and impact is attributed retrospectively under shifting load, traffic mix, or capacity. R2 turns these delivery mechanisms into causal instrumentation. Key challenges: \textit{safe experiments and identification} (designing canary/A--B and quasi-experimental designs that make a change's effect identifiable); \textit{uncertainty tolerance} (propagating uncertainty into go/no-go decisions, e.g., ``promote only if the upper bound on error-rate increase stays within the SLO budget''); \textit{causal impact of configuration/deployment changes} (treating rollouts and mitigations as explicit interventions). If met, R2 yields \textbf{intervention planning} (Fig.~\ref{fig:roadmap}): evidence-backed sequencing of changes to optimize performance, reliability, and cost with quantified uncertainty.

\textbf{Route 3 (R3): Counterfactual assurance (CRL-3: \textbf{\textit{explain \& assure}}).}
Engineers already use counterfactual reasoning in debugging, regression analysis, commit bisection (e.g., \texttt{git bisect}), and postmortems (``what would have prevented this?''). Research also offers program-analysis and testing techniques exploiting counterfactuals \cite{kucuk2021,Baah2010,Giamattei24,Clark23}. However, these methods are often local (single component/execution) or qualitative, and rarely scale to distributed, evolving systems. R3 targets three challenges: \textit{counterfactual RCA at scale} (using traces, topology, and intervention histories to evaluate alternative actions), \textit{causal testing as intervention design} (varying causal factors to expose failures or maximize coverage), and \textit{standard refutation checks} (placebos, sensitivity to adjustment sets, stability across releases \cite{PyWhy}). If met, R3 yields \textbf{counterfactual engineering} (Fig.~\ref{fig:roadmap}): evidence-backed prevention planning guided by credible answers to ``what would have happened under this alternative?''

\textbf{Route 4 (R4): Governance \& alignment (CRL-4/5: \textbf{\textit{justify \& certify}}).}
Organizations already face governance demands -- safety cases, compliance, audits, model risk management -- yet under continuous change, assurance arguments often rest on checklists, correlational evidence, or narrowly scoped tests that do not directly support interventional or counterfactual claims (e.g., ``\textit{this mitigation reduces harm}'' or ``\textit{this policy change is fair under deployment conditions}''). R4 makes causal reasoning usable for assurance and accountability: \textit{auditable causal claims for safety/fairness} (assumptions, evidence, and uncertainty reviewable by stakeholders/regulators), \textit{human-in-the-loop causal modeling} (declaring/revising required or forbidden links, inspecting claim justification), and \textit{causally constrained LLM agents} (actions/explanations respecting admissible interventions, avoiding overclaiming when identification fails). If met, R4 culminates in \textbf{causal copilots}: governed assistants for action and explanation, grounded in causal assumptions, uncertainty, and audit requirements.

\textit{Near-term research bets.}
We distill six research priorities (next 3--5 years), each aligned with a route:
\begin{itemize}[leftmargin=*]
\item \textbf{(R1) Causal variables and models from SE artifacts.} Extract candidate variables and causal relations from logs, traces, configurations, code, and natural-language artifacts. Early attempts exist for causal learning from unstructured data \cite{Liu24_COAT}.
\item \textbf{(R1) Identification under evolution.} Develop version-aware identification and transportability methods that remain valid across releases and changing environments.
\item \textbf{(R2) Safe intervention.} Link effect estimation with staged releases (canaries/A--B tests), enforce uncertainty-aware guardrails (e.g., stop/rollback when worst-case impact is unacceptable).
\item \textbf{(R3) Counterfactual RCA at scale.} Combine causal models with distributed traces and intervention histories to answer counterfactual incident queries and quantify alternative explanations.
\item \textbf{(R3) Causal testing and assurance.} Treat test generation as intervention design; make sensitivity/refutation checks routine to avoid fragile causal claims.
\item \textbf{(R4) Causal copilots for engineers.} Couple LLM agents with causal models as context and constraints, ensuring generated hypotheses and actions are causally valid and uncertainty-aware.
\end{itemize}

\section{Evaluation plan: How to measure progress}
CSE needs shared evaluation targets so different approaches can be compared on common ground. We propose three benchmark families, each mirroring a common engineering task:

\begin{itemize}[leftmargin=*,itemsep=2pt]
\item \textbf{Intervention-effect benchmarks (``what changed because we did $X$?'').}
Curated datasets with known changes (e.g., caching, timeouts, releases) and \emph{ground-truth} impacts (from controlled experiments, replay, or human-assigned labels). These test whether methods can estimate causal effects and whether their \emph{uncertainty} is well calibrated (i.e., confident only when warranted).

\item \textbf{Counterfactual incident benchmarks (``what would have happened if we had acted differently?'').}
Incident datasets including versioned traces/logs, intervention timelines (deployments, rollbacks, mitigations), and outcomes evaluate counterfactual RCA: can a method answer ``would the outage have been avoided if feature flag $F$ had stayed off?'' and produce explanations aligned with known incident narratives.

\item \textbf{Causal testing benchmarks (``which factors \emph{cause} failures, not just correlate with them?'').}
Scenario spaces where causal factors are known or controllable (e.g., simulators, fault-injection environments, replicable microservice workloads). They evaluate whether test generation can identify causal factors and find high-impact failures efficiently, trading effectiveness against intervention cost (time, compute, risk).
\end{itemize}

\textit{Offline and online evaluation.}
Offline benchmarks are necessary but not sufficient. Progress should also be demonstrated in controlled online settings (e.g., canary deployments) using metrics like time-to-root-cause, avoided regressions, and decision confidence.

\textit{Reporting standards and refutation.}
CSE results should report (i)~effect estimates plus uncertainty, (ii)~\emph{sensitivity} and \emph{refutation} analyses, and (iii)~failure modes when assumptions break. \emph{Refutation} means actively testing a causal claim by trying to falsify it --- e.g., checking whether the estimated effect disappears under placebo interventions or changes drastically under reasonable alternative adjustment sets \cite{PyWhy}. A method reporting ``I cannot support this claim under the available data'' is preferable to one producing a confident but fragile attribution.

\section{Conclusion}
Causal Software Engineering re-frames SE as intervention-centric decision-making over systems that evolve in complex environments. 
By making causal assumptions explicit and treating changes as interventions, CSE can enable the next generation of trustworthy assistants, counterfactual testing and debugging, and evidence-backed assurance. We hope this vision catalyzes shared datasets, common evaluation practices, and SE-specific causal tooling that together turn causal reasoning into routine engineering practice.

\section*{Acknowledgment}
This work received funding from the DIETI COSMIC project.

\clearpage
\bibliographystyle{ACM-Reference-Format}
\bibliography{fse26-ivr-cse}

\end{document}